\documentclass{article}

\begin{document}

\def\NA{\ensuremath{N_{1}}}

\def\NB{\ensuremath{N_{2}}}

\title{WAVELENGTH DOUBLING BIFURCATIONS IN A REACTION DIFFUSION SYSTEM}
\author{ Deepak Kar\footnote{E-Mail deepak\_kar@lycos.com}\\ Department of Physics \\Jadavpur University \\Kolkata 700032\\
         India \and J.K.Bhattacharjee\footnote{E-Mail tpjkb@mahendra.iacs.res.in} \\Department of Theoretical Physics\\ Indian
         Institute for Cultivation of Science\\ Kolkata 700032 \\ India }
\maketitle
\begin{abstract} In a two species reaction diffusion system,we show that it
is possible to generate a set of wavelength doubling bifuractions leading to
spatially chaotic state.The wavelength doubling bifurcations are preceded
by a symmetry breaking transition which acts as a precursor.
\end{abstract}
\pagebreak
Reaction Diffusion systems exhibit a rich variety of spatial patterns.
The simplest such system consists of two species,one of which is autocatalytic and the other a fast diffusing antagonist.The interplay
between the two gives rise to intersting Turing patterns \cite{1}.A popular
model is that of Gierer and Meinhardt\cite{2,3,4}.It has the form
\[\dot{A}=D\nabla^{2}A+A^{2}/B-A+\sigma \]
\[\dot{B}=\nabla^{2}B+\mu(A^{2}-B^{2})\]
Here A and B are the two interacting species.The three parameters in
the problem are D, the ratio of the diffusion coefficient of species A
and B, $\sigma$, the rate of production of A and $\mu$ the reaction rate
for the production of B.The homogenous steady state
$A^{2}=B=(1+\sigma)^{2}$ is unstable to a patterned state along the
boundary $\mu_{c}D=[(2 /(1+\sigma))^{1/2}-1]^{2}$
and for $\mu<\mu_{c}$ for a given D. The homogenous steady state
is unstable to a homogenous but temporal periodic state for
$\mu<(1-\sigma) /(1+\sigma)$.In the region close to $D\simeq0$ and
$\mu\simeq1$ there is a chaotic state where the Hopf and Turing
regions overlap\cite{5,6}.Steady but spatially chaotic states are however
not known in the Gierer-Meinhardt model.On the other hand spatially
chaotic states \cite{7} are known to arise in the coupled map lattices\cite{8,9,10}
which have recieved a lot of attention in the last decade.In this work,we
have taken a two species model which is somewhat simpler than the Geirer-
Meinhardt model above and explicitly shown the existance of a spatially
chaotic state obtained by a process of wavelength doubling. It is interesting to note that the wavelength doubling is preceded by a symmetry breaking bifurcation similar to what occurs in a damped driven anharmonic
harmonic oscillator\cite{11,12}.

We consider a two species model where one of the species can be self saturating or self catalysing with a similar option for the other.The
species interact via a linear cross coupling.Each species can help or hinder the growth of the
other.The model for the populations $N_{1},
N_{2}$ of the two species is
\begin{equation}\dot{\NA}=D\nabla^{2}\NA+a_{1}\NA(1-\NB^{2})+b_{2}\NA \end{equation}
\begin{equation}\dot{\NB}=\nabla^{2}\NB+a_{2}\NB(1-\NA^{2})+b_{2}\NA\end{equation}

\pagebreak
A spatially homogenous solution is one where both species die
out i.e $\NA=\NB=0$. The linear stability of the state in the absence
of diffusion term shows that this will be a stable state of affairs if
$a_{1}+a_{2}<0$ and $a_{1}a_{2}-b_{1}b_{2}>0$. We will consider $a_{1}$,
$a_{2}$ , $b_{1}$ and $b_{2}$ to always satisfy thse constraints.
Consequently our focus will be on an instability brought about by
the diffusion terms.The linear stability of $\NA=\NB=0$ against a
perturbation $\delta N_{1,2}=\delta_{1,2}e^{ikx}$ leads to the linearized
equations
\begin{equation}\dot{\delta_{1}}=(-Dk^{2}+a_{1})\delta_{1}+b_{1}\delta_{2}\end{equation}
\begin{equation}\dot{\delta_{2}}=(-k^{2}+a_{2})\delta_{2}+b_{2}\delta_{1}\end{equation}
The growth rate p of the perturbation $\delta_{1,2}$ can be found
from
\begin{equation}Det\pmatrix{a_{1}-Dk^{2}-p & -b_{1}\cr -b_{2} & a_{2}-k^{2}-p\cr}=0\end{equation}
so that
\begin{eqnarray}& 2p &=  a_{1}+a_{2}-(D+1)k^{2}\pm \nonumber \\
                   & & \sqrt{(a_{1}+a_{2}-(D+1)k^{2})^{2}+
 4(a_{1}a_{2}-b_{1}b_{2}-k^{2}(Da_{2}+a_{1})+Dk^{4})} \end{eqnarray}
and instability can occur if
\begin{equation}a_{1}a_{2}-b_{1}b_{2}-k^{2}(Da_{2}+a_{1})+Dk^{4}<0\end{equation}
The extremum of the function in the left hand side of Eq(7) occurs at
\begin{equation}k^{2}=k_{0}^{2}={1\over2}(a_{2}+{a_{1}\over D}) \end{equation}
and enforcing Eq(7) at this wavenumber, we need
\begin{equation}(a_{2}D+a_{1})^{2}>4D(a_{1}a_{2}-b_{1}b_{2})\end{equation}
This condition can be exactly satisfied and this implies that
in the presence of diffusion ,the previously stable state can
be destabilized in favour of a spatially periodic state.

If we want to explore the stationary spatially periodic state
,then we need to write
\begin{equation}N_{1,2}=A_{1,2}cosk_{0}x\end{equation}
and find the coefficients $A_{1,2}$.This can be done by using the
method of harmonic balance on Eq(1,2).We note immediately
that terms of the form $cos3k_{0}x$ will be generated.For the moment we ignore them and straightforward algebra leads to
\begin{equation}(a_{1}-Dk_{0}^{2})A_{1}-{3\over4}a_{1}A_{1}^{3}+b_{1}A_{2}=0\end{equation}
\begin{equation}(a_{2}-k_{0}^{2})A_{2}-{3\over4}a_{2}A_{2}^{3}+b_{2}A_{1}=0\end{equation}

We have verified that the cubic equation for $A_{1}^{2}$ that
results from the above gives a unique answer i.e that there is
only one positive root for a large range of values of the diffusion
constant D.

Returning now to the terms dropped in our harmonic balance we note that
$cos^{3}k_{0}x$ produces a $cos3k_{0}x$ term which implies that the response will have a third harmonic as well, i.e the solution will be
\begin{equation}N_{1,2}=A_{1,2}cosk_{0}x+B_{1,2}cos3k_{0}x+\ldots \end{equation}
It is easy to see that $N_{1,2}^{3}$ will now yield terms like
$A_{1,2}^{2}cos^{2}k_{0}xB_{1,2}cos3k_{0}x$ which will produce
the fifth harmonic.Consequently in the expansion above all the odd
harmonics will be present but not the even harmonics.This immediately
means that the solution has a symmetry-namely if $x\rightarrow x+(2m+1){\pi\over k_{0}}$ then $N(x+(2m+1){\pi\over k_{0}})=-N(x)$.
So long as a solution has this symmetry there cannot be a wavelength
doubling bifurcation.

We then ask the question if the above symmetry can be broken by the introduction of a $cos2k_{0}x$ term.

We try the stability of the solution of odd harmonics represented by Eq(13) by adding a perturbation $\delta N_{1,2}=\delta_{1,2}e^{pt}cos2k_{0}x$.
Straightforward algebra leads to the following qudratic equation for p,
\begin{equation}p^{2}-(\alpha_{1}+\alpha_{2}-4k_{0}^{2}(D+1))p-4(\alpha_{1}\alpha_{2}
-b_{1}b_{2})+16k_{0}^{2}(D\alpha_{2}+\alpha_{1})-16Dk_{0}^{4}=0\end{equation}
where
\begin{equation}\alpha_{1,2}=a_{1,2}[1-{3\over2}(A_{1,2}^{2}+B_{1,2}^{2}+A_{1,2}B_{1,2})]\end{equation}

With $\alpha_{1}+\alpha_{2}<0$ and $\alpha_{1}\alpha_{2}-b_{1}b_{2}>0$ it is clear that p
can only be negative.In the presence of diffusion,instability will set in if
\begin{equation}4k_{0}^{2}(D\alpha_{2}+\alpha_{1})-16Dk_{0}^{4}>\alpha_{1}\alpha_{2}
-b_{1}b_{2}\end{equation}
We can tune the cross coupling coefficient or the self coupling to ensure
that the above condition is satisfied.The broken-symmetry solution will
look like
\begin{equation}N_{1,2}=A_{1,2}cosk_{0}x+B_{1,2}cos3k_{0}x+C_{1,2}cos2k_{0}x+\ldots\end{equation}

Just like the coefficients $A_{1,2}$ and $B_{1,2}$ ,the coefficients $C_{1,2}$ can be found from harmonic balance to be given by
\begin{equation}(a_{1}-4k_{0}^{2}D)C_{1}-a_{1}({3A_{1}^{2}\over 2}+{3B_{1}^{2}\over 2}
 +{3A_{1}B_{1}\over 2}+{3C_{1}^{2}\over 4})c_{1}=-b_{1}C_{2}\end{equation}
\begin{equation}(a_{2}-4k_{0}^{2})C_{2}-a_{2}({3A_{1}^{2}\over 2}+{3B_{1}^{2}\over 2}
 +{3A_{1}B_{1}\over 2}+{3C_{2}^{2}\over 4})c_{2}=-b_{2}C_{1}\end{equation}

The symmetry breaking solution of Eq(1,2) now allows wavelength doubling to
take place.To see this we write
\begin{equation}N_{1,2}=A_{1,2}cosk_{0}x+B_{1,2}cos3k_{0}x+C_{1,2}cos2k_{0}x+\delta_{1,2}cos{kx\over 2}\end{equation}

Linearizing Eq(1,2) in $\delta_{1,2}$ ,we find
\begin{equation}\dot{\delta_{1}}=-{k^{2}\over 4}\delta_{1}+a_{1}\delta_{1}+b_{1}\delta_{2}-3a_{1}(A_{1}+B_{1})C_{1}\delta_{1}\end{equation}
\begin{equation}\dot{\delta_{2}}=-{k^{2}\over 4}\delta_{2}+a_{2}\delta_{2}+b_{2}\delta_{1}-3a_{2}(A_{2}+B_{2})C_{2}\delta_{2}\end{equation}
By defining
\begin{equation}\beta_{1,2}=a_{1,2}(1-3(A_{1,2}+B_{1,2})C_{1,2})\end{equation}
and we can write Eqs(21) and Eqs(22) as
\begin{equation}\dot{\delta_{1}}=(\beta_{1}-{k^{2}\over 4})\delta_{1}+b_{1}\delta_{2}\end{equation}
\begin{equation}\dot{\delta_{2}}=b_{2}\delta_{1}+(\beta_{2}-{k^{2}\over 4})\delta_{2}\end{equation}

The solution for $\delta_{1,2}$ will be of the form $e^{pt}$.
The condition for genration of $cos{k_{0}x\over 2}$ - the wavelength doubled state-can now be easily obtained in a form similar to Eq(16).Our
new state with anharmonic response can be written as
\begin{equation}N_{1,2}=A_{1,2}cosk_{0}x+B_{1,2}cos3k_{0}x+C_{1,2}cos2k_{0}x+S_{1,2}cos{k_{0}x\over 2}+\ldots\end{equation}
where the amplitudes $S_{1,2}$ can be found by using harmonic balance.
To generate the next wavelength doubling one tries
\begin{equation}N_{1,2}=A_{1,2}cosk_{0}x+B_{1,2}cos3k_{0}x+C_{1,2}cos2k_{0}x+S_{1,2}cos{k_{0}x\over 2}+\delta_{1,2}cos{k_{0}x \over 4}\end{equation}
and linearizes Eq(1,2) in $\delta_{1,2}$.The resulting equations are
\begin{equation}\dot{\delta_{1}}=(a_{1}-{Dk_{0}^{2}\over 16})\delta_{1}+b_{1}\delta_{2}
-{3 \over 4}a_{1}A_{1}S_{1}\delta_{1}\end{equation}
\begin{equation}\dot{\delta_{2}}=(a_{2}-{Dk_{0}^{2}\over 16})\delta_{2}+b_{2}\delta_{1}
-{3 \over 4}a_{2}A_{2}S_{2}\delta_{2}\end{equation}

As we have done several times before,we can determine the condition for growth rate to be positive and thus the period quadrupled state is
generated.The new state has the structure
\begin{equation}N_{1,2}=A_{1,2}cosk_{0}x+B_{1,2}cos3k_{0}x+C_{1,2}cos2k_{0}x+S_{1,2}cos{k_{0}x\over 2}+S'_{1,2}cos{k_{0}x \over 4}\end{equation}
The sequence is now clear.We can add a perturbation $\delta_{1,2}cos{k_{0}x\over 8}$ and the term $S_{1,2}S'_{1,2}$ will trigger
the instability that will lead to the generation of this longer wavelength
state.A continous sequence of doubling can now take place leading to a state with no definite wavelengths.

Thus we see that in a two-species
model with one of the species autocatalytic and the other showing
self saturation,we can have a spatially chaotic state attained through a succession of wavelength doubling bifurcations.It is interesting that the 
wavelength doubling bifurcations has to be preceeded by a symmetry breaking bifurcation.

\section*{Acknowledgements}
One of the authors (Deepak Kar) gratefully acknowledges financial support
from Indian Academy of Sciemces in the form of a Summer Research Fellowship.

\end{document}